\documentclass[a4paper,11pt]{article}
\pdfoutput=1
 
\usepackage{jinstpub} 

\usepackage{graphicx}
\usepackage{color}
\usepackage{mathtools}
\usepackage[left]{lineno}
\usepackage{subfigure}

\newcommand{\BX}{LP2I Bordeaux, Universit\'{e} de Bordeaux, CNRS/IN2P3, F-33175 Gradignan, France}
\newcommand{\CEA}{IRFU, CEA, Universit\'{e} Paris-Saclay, F-91191 Gif-sur-Yvette, France}
\newcommand{\CPPM}{ CPPM, Universit\'{e} d'Aix-Marseille, CNRS/IN2P3, F-13288 Marseille, France}
\newcommand{\LSM}{ LSM, CNRS/IN2P3, Universit\'{e} Grenoble-Alpes, Modane, France}
\newcommand{\SUBATECH}{ SUBATECH, IMT-Atlantique, Universit\'{e} de Nantes, CNRS-IN2P3, France}
\newcommand{\BIRMINGHAM}{ School of Physics and Astronomy, University of Birmingham, B15 2TT, United Kingdom}

\title{Performance of a spherical high pressure gas TPC for neutrino magnetic moment measurement}

\author[a]{R.~Bouet,}
\author[b]{J.~Busto,}
\author[a]{V.~Cecchini,}
\author[a]{C.~Cerna,}
\author[a]{P.~Charpentier,}
\author[c]{A.~Dastgheibi-Fard,}
\author[a]{F.~Druillole,}
\author[a,1]{C.~Jollet,\note{Corresponding author.}}
\author[a]{P.~Hellmuth,}
\author[d]{I.~Katsioulas,}
\author[d,e]{P.~Knights,}
\author[e]{I.~Giomataris,}
\author[e]{M.~Gros,}
\author[f]{P.~Lautridou,}
\author[a]{A.~Meregaglia,}
\author[e] {X.~F.~Navick,}
\author[d]{T.~Neep,}
\author[d]{K.~Nikolopoulos,}
\author[a]{F.~Perrot,}
\author[a]{F.~Piquemal,}
\author[a]{M.~Roche,}
\author[a]{B.~Thomas,}
\author[d]{R.~Ward}

\affiliation[a]{\BX}
\affiliation[b]{\CPPM}
\affiliation[c]{\LSM}
\affiliation[d]{\BIRMINGHAM}
\affiliation[e]{\CEA}
\affiliation[f]{\SUBATECH}

\emailAdd{cecile.jollet@cern.ch}

\abstract{The measurement of neutrino magnetic moment larger than $10^{-19}\mu_B$ would be a clear signature of physics beyond the standard model other than the existence of massive Dirac neutrinos. The use of a spherical proportional counter detector filled with gas at 40 bar located near a nuclear reactor would be a simple way to perform such a measurement exploiting the developments made on such a technology for the search of dark matter and neutrinoless double beta decay. Different targets can be used just by replacing the gas: xenon, CF$_4$ and argon were compared and the sensitivity in one year of data taking could reach the level of $4.3 \times 10^{-12} \mu_B$, $6.5 \times 10^{-12} \mu_B$, and  $8.5 \times 10^{-12} \mu_B$, respectively.
}

\begin{document} 

\maketitle
\flushbottom


\section{Introduction}

The observation of a non-zero value of the neutrino magnetic moment (NMM) $\mu_\nu$ is an important signature of physics beyond the Standard Model (SM)~\cite{Bell:2005kz}.
The Minimally extended Standard Model~\cite{Asaka:2005pn} (MSM) predicts for a massive Dirac neutrino a magnetic moment of $3.2 \times10^{-19} (\frac{m_i}{eV}) \mu_B$ where $m_i$ is the neutrino mass and $\mu_B$ is the Bohr magneton. Considering the best existing constraints on the neutrino masses of 0.8~eV~\cite{KATRIN:2021uub} the predicted neutrino magnetic moment is at the level of $4.0 \times10^{-19}\mu_B$,  which is several orders of magnitude smaller than the current experimental sensitivity.
The possible observation of a neutrino magnetic moment much larger than $10^{-19}\mu_B$,  is particularly interesting since it would highlight beyond SM physics other than massive Dirac neutrinos. There are indeed several extensions beyond the MSM for which the neutrino magnetic moment, for a Majorana neutrino, could be at the level of $10^{-12}\mu_B$~\cite{Broggini:2012df,Studenikin:2016ykv,Gorchtein:2006na}. In general, the coupling of neutrinos with photons is a generic consequence of the finite neutrino masses, and is one of the important intrinsic neutrino properties to explore. 

 Experimental constraints on the neutrino magnetic moment have been obtained from reactor neutrino experiments, solar neutrinos, and astrophysical measurements. The most established and sensitive method for the experimental investigation of neutrino electromagnetic properties is provided by direct laboratory measurements of anti-neutrino electron scattering at nuclear reactors. The signature is an excess of recoil electron events over those due to standard model and other background processes, which exhibit a 1/T spectral dependence, where T is the electron recoil kinetic energy.\\ 
The Borexino experiment obtained an upper limit on a neutrino magnetic moment of $\mu_\nu < 2.8 \times 10^{-11} \mu_B$~\cite{Borexino:2017fbd}.
Additional stringent upper bounds on the neutrino magnetic moment have been obtained by the GEMMA collaboration~\cite{Beda:2013mta}: with an energy threshold of about 2.8 keV  for electron recoil and 1.5~kg germanium detector, the neutrino magnetic moment is bounded by $\mu_\nu < 2.9 \times 10^{-11} \mu_B$. The use of the Time Projection Chamber (TPC) technique was also employed in such a search yielding a limit of $\mu_\nu < 9 \times 10^{-11} \mu_B$  by the MUNU collaboration with a TPC filled with 10 kg  of CF$_4$ and an energy threshold of 700~keV~\cite{Daraktchieva:2008gb}. Very recently the XENON1T collaboration released results on solar neutrino analysis obtaining a limit of $6.3 \times 10^{-12} \mu_B$~\cite{XENONCollaboration:2022kmb}.
 
To search for such a rare process the ideal detector has a large target mass, a low radioactivity background, a low energy threshold, and a good energy resolution. Therefore, the use of a spherical high pressure gas TPC, or spherical proportional counter (SPC), could be a valid option for a detector aiming at NMM observation. Such a detector, thanks to its features, including an extremely low radioactivity background, the sensitivity to single electron signals, and the excellent energy resolution, is used today for the direct search of dark matter by the NEWS-G collaboration~\cite{Gerbier:2014jwa,Arnaud:2017bjh} and it is under study for the search of neutrinoless double beta decay ($\beta\beta0\nu$) by the R2D2 collaboration~\cite{Meregaglia:2017nhx,Bouet:2020lbp}.

In this paper the performance of the SPC detector, filled with different gases, is investigated, and for the most promising configuration a full simulation of all the different backgrounds is carried out. Assuming operation of the detector at a distance of about 10~m from a research or commercial nuclear reactor, an upper limit between $4.3 \times 10^{-12} \mu_B$ and $8.5 \times 10^{-12} \mu_B$ could be obtained  (depending on the gas) on NMM. Such a limit would be of the order of the recently published XENON1T one but obtained detecting reactor neutrinos instead of solar ones.\\
It has to be pointed out that despite the efforts made to have a Monte-Carlo (MC) simulation of the experimental setup as realistic as possible, several mechanical features such as feedthroughs and supports of the different vessels were not accounted for, with a possible impact on the final sensitivity. 


\section{Flux and cross sections}
\label{sec:cross}
To estimate the integral and the energy spectrum of the observed neutrinos, when locating the SPC detector near a nuclear reactor, the two key ingredients are the initial neutrino flux and the interaction cross section.
The neutrino flux depends on the type of reactor (research or commercial one) and in particular on the different fraction of the fissile elements: a commercial reactor will have a smaller fraction of U$^{235}$ compared to a research reactor. Such a difference, however, introduces second order effects, and in the present study we assumed the Mueller et al. parameterization~\cite{Mueller:2011nm} for energies above 2~MeV and the Kopeikin one~\cite{Kopeikin:2012zz} below 2~MeV.
The absolute normalization was fixed using the NUMU measurement at Bugey, and a flux of $10^{13}$ anti-neutrinos cm$^{-2}$ s$^{-1}$ was assumed, corresponding to a distance from the reactor of 18~m.
%

The process considered to observe anti-neutrinos is elastic scattering on electrons. The electro-weak cross section depends on the anti-neutrino energy $E$ and on the energy of the outgoing electron $T$ and in case of massless SM neutrino can be written as~\cite{tHooft:1971ucy}:

\begin{equation}
\left (\frac{d\sigma}{dT} \right )= \frac{G_F^2 m_e}{2\pi} \left ( (g_V-g_A)^2 + (g_V+g_A)^2 (1-\frac{T}{E})^2 -(g_V^2-g_A^2) \frac{m_e T}{E^2} \right )\\
\label{eq:eq1}
\end{equation}
where $m_e$ is the electron mass, $G_F$ the Fermi coupling constant,
\begin{equation}
\frac{G_F^2 m_e}{2\pi}=\frac{4.305 \times10^{-42} cm^2}{1 GeV},\\
\end{equation}
\begin{equation}
g_V=1 - 0.5 + 2 \sin^2 \theta_w,\\
\end{equation}
\begin{equation}
g_A=1 - 0.5.\\
\end{equation}
The weak mixing angle $\sin^2\theta_w$ is assumed to be 0.23857 as taken from the Particle Data Group~\cite{ParticleDataGroup:2018ovx}.
For each neutrino energy, the total cross section is computed integrating over $dT$ between the threshold of 20~eV and the maximal kinematically allowed energy $T_{max}$ with:
\begin{equation}
T_\text{max}=\frac{E}{1+\frac{m_e}{2E}}.
\label{eq:eq2}
\end{equation}
The additional cross section due to NMM can be written as:
\begin{equation}
\label{eq:eq3}
\left (\frac{d\sigma}{dT} \right ) = \pi r_e^2 \left(\frac{\mu_\nu}{\mu_B} \right)^2 \left(\frac{1}{T}-\frac{1}{E} \right) 
\end{equation}
where $r_e$ is the electron radius and $\mu_\nu$ the NMM in units of $\mu_B$.\\

\begin{figure} [t]
\begin{center}
\subfigure[\label{fig:cs1}]{\includegraphics[height=5 cm]{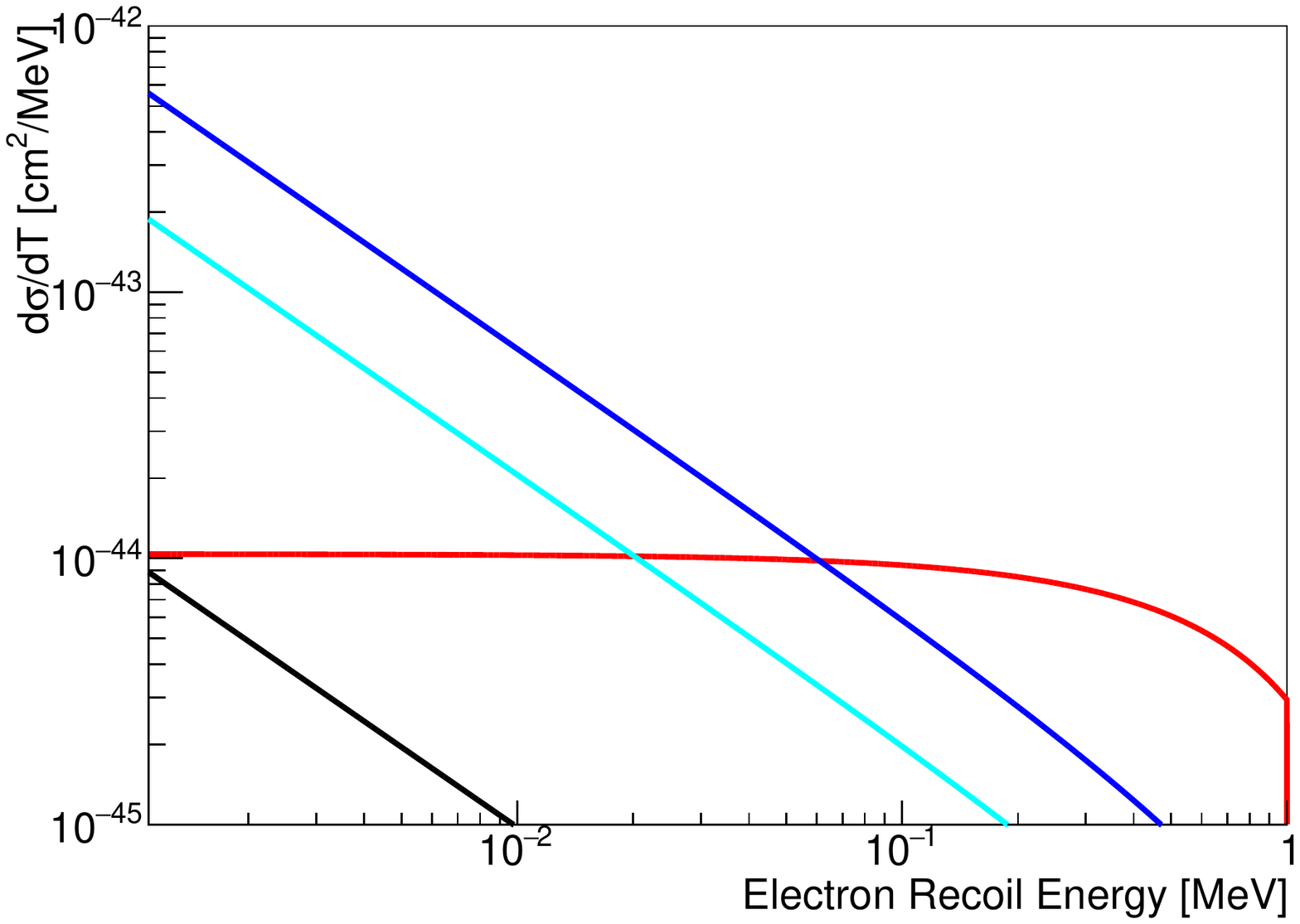}}
\subfigure[\label{fig:cs2}]{\includegraphics[height=5 cm]{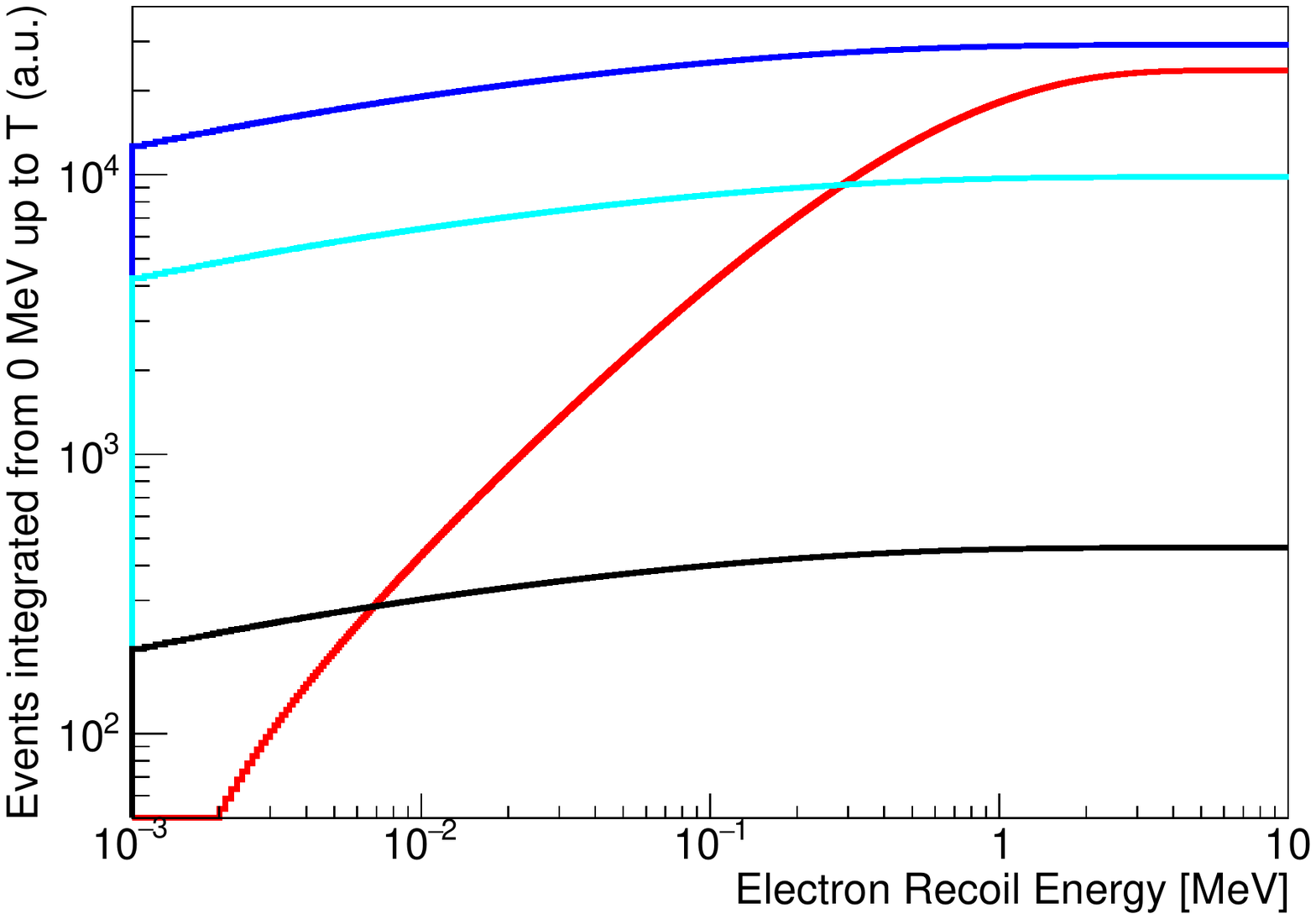}}
\caption{{\it Neutrino-electron scattering cross section~\subref{fig:cs1} and number of integrated events~\subref{fig:cs2} for electroweak physics (red line) and NMM physics assuming $\mu_\nu=5\times 10^{-11} \mu_B$ (blue line), $\mu_\nu=2.9\times 10^{-11} \mu_B$ (cyan line), and $\mu_\nu=6.3\times 10^{-12} \mu_B$ (black line).}}
\end{center}
\end{figure}

The total cross section for the electro-weak process and for the NMM one are shown in Fig.~\ref{fig:cs1} for a mean neutrino  energy of 2~MeV and for three different values of $\mu_\nu$. Once the cross section is correctly integrated over the neutrino spectrum and over the electron  recoil energy the number of events can be obtained. Its integral in the range [0~MeV--$T$] as a function of $T$ is shown in Fig.~\ref{fig:cs2}. The number of events is given here in arbitrary units just for comparison between the different processes, since the real normalization depends on the number of electrons and therefore on the mass and type of target used.  It is clear that the NMM contribution is dominant at low energy, and given the dependence of the cross section as $\mu_\nu^2$, the excess of events due to NMM quickly drops when the $\mu_\nu$ gets smaller.

\section{Detector setup}
The SPC detector is a grounded sphere filled with gas at high pressure. At the center of the sphere there is the anode consisting of a few mm diameter sphere at a positive high voltage. The generated electric field is used to drift the electrons, produced by the particles crossing and ionizing the gas, towards the central anode. When the electrons are close enough to the anode they enter the avalanche region and they are collected by the anode itself to form the readout signal. More details on the working principle of the spherical TPC detector can be found in Ref.~\cite{Gerbier:2014jwa,Savvidis:2016wei}.\\
The geometry of the proposed detector is based on the design optimized for the search of neutrinoless double beta decay~\cite{Meregaglia:2017nhx} with some changes needed to reduce the background coming from external neutrons and decays of cosmogenic nuclei. Indeed, in $\beta\beta0\nu$ searches the detector should be located in an underground laboratory, whereas for the search of NMM the detector should be operated near a nuclear reactor: the small overburden results therefore in a much higher cosmic background. The proposed onion-like structure (see Fig.~\ref{fig:detector}) is the following:
\begin{figure} [t]
\begin{center}
\includegraphics[width=0.9\textwidth]{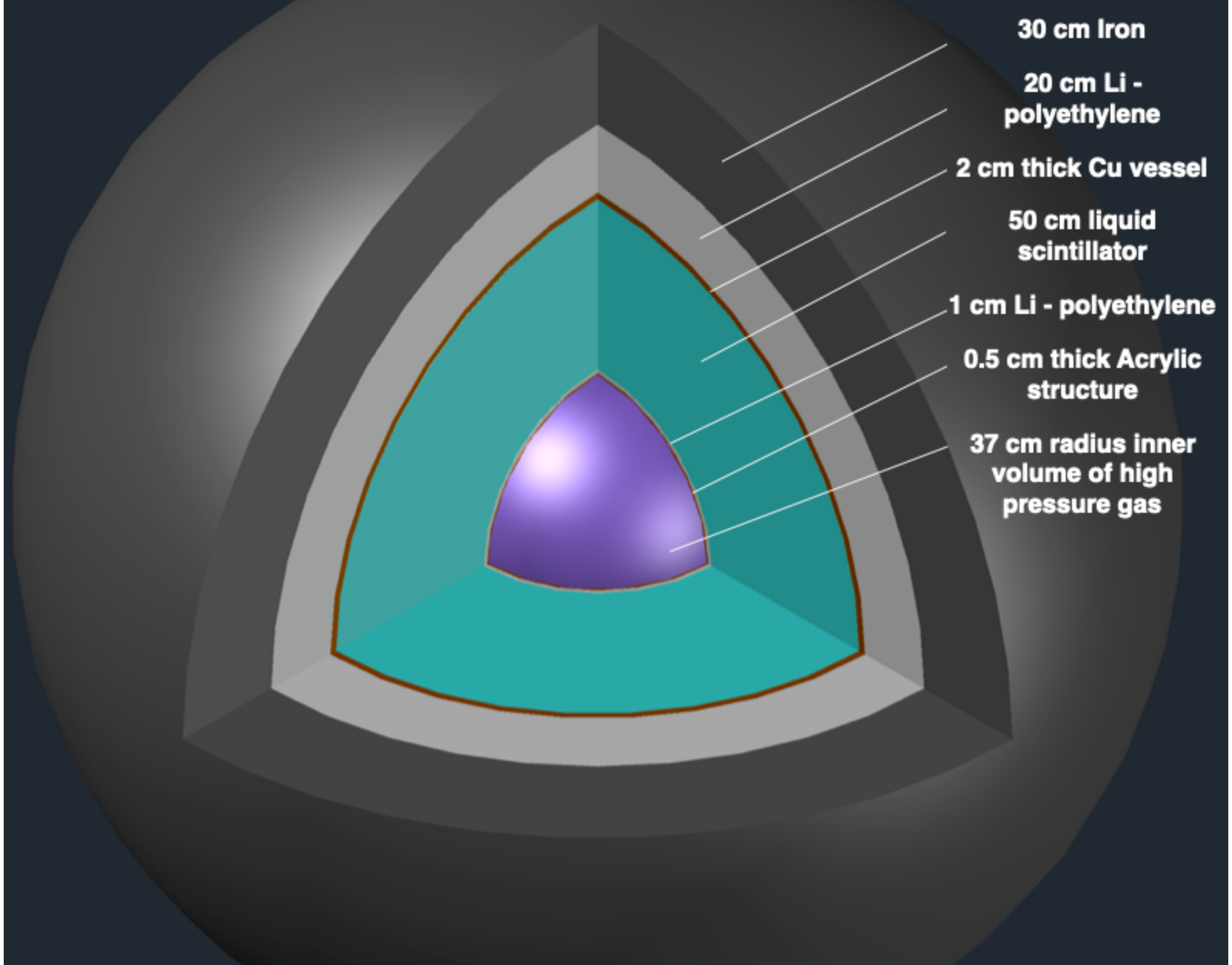}
\caption{{\it Cartoon showing the detector setup.}}
\label{fig:detector}
\end{center}
\end{figure}
\begin{itemize}
\item {\it Gas active volume}\\ Inner sphere filled with gas at high pressure with a radius of 37~cm and instrumented with a central anode. Different gases and pressures were considered  namely CF$_4$, Ar and Xe. The optimal gas pressure is 40~bar (see Sec.~\ref{sec:BG}) corresponding to a total mass of 36.3~kg, 14.3~kg, and 63~kg, respectively.
\item {\it Acrylic vessel}\\ The active volume is contained in an acrylic sphere with a thickness of 0.5~cm. A study carried out by the mechanical service of LP2I Bordeaux showed that such a thickness is enough to assure the mechanical stability of the detector. Indeed, a thickness of only 2.1~mm would be already sufficient if the pressure difference between the gas active volume (inside) and the liquid scintillator (outside) is kept below 0.5~bar. Indeed the difference between top and bottom of the detector due to the hydrostatic pressure of the liquid scintillator is below 0.1 bar, and pressurizing the liquid while filling with gas the inner volume will result in an almost constant pressure difference across the acrylic vessel.
The mechanical study was carried out for a perfectly spherical detector since at this point we only have a conceptual detector and no feedthroughs or internal supporting structure were simulated. It has to be mentioned that copper can not be used (baseline for a $\beta\beta0\nu$ detector~\cite{Meregaglia:2017nhx}) since given the large cosmic muon flux due to the low overburden, the cosmogenic production of cobalt isotopes is extremely important~\cite{Baudis:2015kqa} resulting in a background definitely too high.
\item{\it Li-Polyethylene shielding}\\ A thin layer of 1~cm of Li-doped polyethylene at 7~\% is foreseen to capture thermal neutrons. The advantage of using lithium instead of boron to capture neutrons is the fact that no gammas are emitted in the process reducing possible contributions to the background. 
\item {\it Liquid scintillator volume}\\ The acrylic sphere is contained in a spherical liquid scintillator (LS) volume, assumed to be Linear Alkylbenzene (LAB), with a thickness of 50~cm  which is used as veto volume. The volume would be instrumented with wavelength shifting fibers to collect the scintillation light and transport it outside of the detector where the fibers are connected to photomultiplier tubes. Such a veto provides a rejection of both events coming from the outside, and radioactive events generated in copper, in particular from $^{208}$Tl and cosmogenic Co isotopes. The thickness was optimized in order to be as small as possible without affecting the background reduction. The LS is also used to compensate the gas pressure and avoid a large mechanical stress on the acrylic vessel.
\item {\it Liquid scintillator copper vessel}\\ The vessel is a sphere of a thickness of 2~cm made of copper to minimize the possible background. This vessel is far enough from the active volume to allow copper as building material.
\item {\it Shielding}\\ An external shielding to mainly reduce the neutron induced background is necessary. We assumed, starting from the inside, a layer of 20~cm of  Li-doped polyethylene and 30~cm of iron. Neutrons interacting in the iron layer are either absorbed or yield lower energy neutrons that can be absorbed in the light polyethylene layer. Gammas are also emitted in neutron interactions with iron, but the LS will veto most of them. \end{itemize}

 The present study considered a conceptual detector design. Several realistic detector configurations are currently under study including the possibility to use composite materials withstanding a pressure of 85~bar with only 4~mm thickness~\cite{tank}. In this case there would be no need of liquid scintillator pressurization and the feedthrough and connections would be much easier.


\section{Background studies}
\label{sec:BG}
The different sources of backgrounds were studied thanks to a full Monte-Carlo simulation based on Geant4 toolkit~\cite{Agostinelli:2002hh}. Details on the studied sources of backgrounds, as well as on the signal selection analysis results, are given in the following subsections. Further details on the backgrounds, including their energy distributions before and after selection cuts, can be found in Ref.~\cite{Meregaglia:2017nhx}.

\subsection{Electro-weak signal}
The SM electro-weak signal is an irreducible background when searching for NMM signal.\\
Considering the cross section of the two phenomena (see Fig.~\ref{fig:cs1}) it is clear that at energies of the order of few tens of keV, depending on the value of the NMM, the SM process dominates. For this reason an upper cut is used to define the region of interest (ROI) and reduce the impact of the electro-weak signal (45~keV in our case as explained in Sec.~\ref{sec:signal}).

\subsection{Acrylic vessels radioactivity}
Another irreducible background is due to the radioactive decay chains of the acrylic vessel.\\ 
The contaminations in terms of $^{238}$U and $^{232}$Th were assumed to be 3.7~$\mu$Bq/kg and 1.2~$\mu$Bq/kg, respectively, as measured for the acrylics of the JUNO detector~\cite{Cao:2020zyr}. A factor of 3 improvement is probably still possible however this was not accounted for in the present study.

\subsection{Copper vessels radioactivity}
A full simulation of the  $^{238}$U and $^{232}$Th decay chain in the copper vessel containing the liquid scintillator was carried out, assuming an activity of 10~$\mu$Bq/kg. This is a conservative limit since copper with a bulk radioactivity 10 times better could be found on the market (e.g. from the AURUBIS company~\cite{aurubis}). However, given the fractional impact of such a background (see Tab.~\ref{tab:BG}), a reduction of an order of magnitude on the copper purity would only result in a total background reduction of about 2\%.\\
Radioactivity inside the liquid scintillator was not simulated since it is normally extremely pure, as already demonstrated by experiments such as KamLAND~\cite{Suekane:2004ny} or SNO+~\cite{SNO:2021xpa}, yielding therefore a negligible contribution to the total background.

\subsection{Radon}
Radon ($^{222}$Rn) is an important source of background for all experiments searching for rare events. Despite all the efforts made to build detectors in clean rooms and the use purification systems for fluids, radon is often the limiting factor for the experimental sensitivity.\\
In the proposed detector $^{222}$Rn is expected to be present in the liquid scintillator and in the gas itself. Based on the SuperNEMO experience~\cite{Hodak:2019qty}, with a recirculation of the gas using a J-Trap, a contamination at the level of 20~$\mu$Bq/m$^3$ could be achieved, corresponding to 5~$\mu$Bq/kg. With such a radon contamination the background due to liquid scintillator is negligible, whereas the contribution due to the presence of Rn in the gas accounts for a fraction of background between 15\% and 30\% depending on the gas.

A second effect of radon is that during the detector production it deposits on surfaces. Following its decay chain, it results in surface contamination of $^{210}$Pb. Experimental measurements at Modane underground laboratory (LSM) in France with the SEDINE SPC detector~\cite{Fard:2015pla} showed that a contamination at the level of 0.2~$\mu$Bq/cm$^2$ could be achieved. 
Considering the surface of the liquid scintillator copper vessel, the impact of such a background depends on the gas and its pressure, accounting for a larger fraction of the total background in light gases and at low pressure. This is mainly due to the selection cut which rejects events contained in the outermost cm of the detector: the higher the pressure the higher the probability of containing those events in the outer detector skin, and the same dependence holds according to the gas mass. This background contribution could be reduced increasing the radial selection cut at the price of reducing the active volume. Although the obtained background value is not a show stopper for the proposed experiment, further efforts are ongoing for a cleaner detector, and better results are expected from the NEWS-G detector~\cite{Gerbier:2014jwa}.

\subsection{Neutrons}

Neutrons represent an important source of background and a detailed simulation is complicated since it strongly depends on the detector location, shielding and overburden.\\
The first source of neutron is the reactor itself, however based on previous experiments experience such as STEREO~\cite{Allemandou:2018vwb} or CONUS~\cite{Hakenmuller:2019ecb} it seems that such a source of background can be neglected. For STEREO this is possible thanks to the specific neutron shielding, whereas CONUS, thanks to a comparison between measurements with reactor on and off, finds that the neutron background correlated with reactor is negligible.\\
Therefore, only neutrons induced by cosmic rays are considered in the background study. The spectra measured in Ref.~\cite{1369506} was used as input in the full Monte-Carlo simulation with an additional overburden of 3~m of concrete or 7.2~m water equivalent (m.w.e.). The overburden depends of course on the detector location but 3~m of concrete is a conservative estimate for several possible detector sites such as the Institut Laue-Langevin (ILL) (15 m.w.e. for STEREO) of the KBR Brokdorf (24 m.w.e. for CONUS).\\
The neutrons are generated outside the detector shielding and the different processes including radiative absorption are simulated by the Geant4 toolkit. The events passing the selection cuts, rescaled to one year of data taking, account for about 10\% to 12\% of the total background depending on the gas.

An additional contribution to the neutron background comes from fast neutrons produced by cosmic muons. A study of CONUS showed that the largest contribution comes from neutrons produced directly in the experimental shielding~\cite{Hakenmuller:2019ecb}. This of course depends on the detector geometry and on the experimental site, however assuming their measured flux, and simulating neutrons of 1~MeV on the inner side of the shielding, a number of expected events smaller than one per year was obtained. Such a background contribution was therefore neglected in our study.

\subsection{Signal selection and background rejection results}
\label{sec:signal}
\begin{figure} [p]
\begin{center}
\subfigure[\label{fig:cut1}]{\includegraphics[height=6.5 cm]{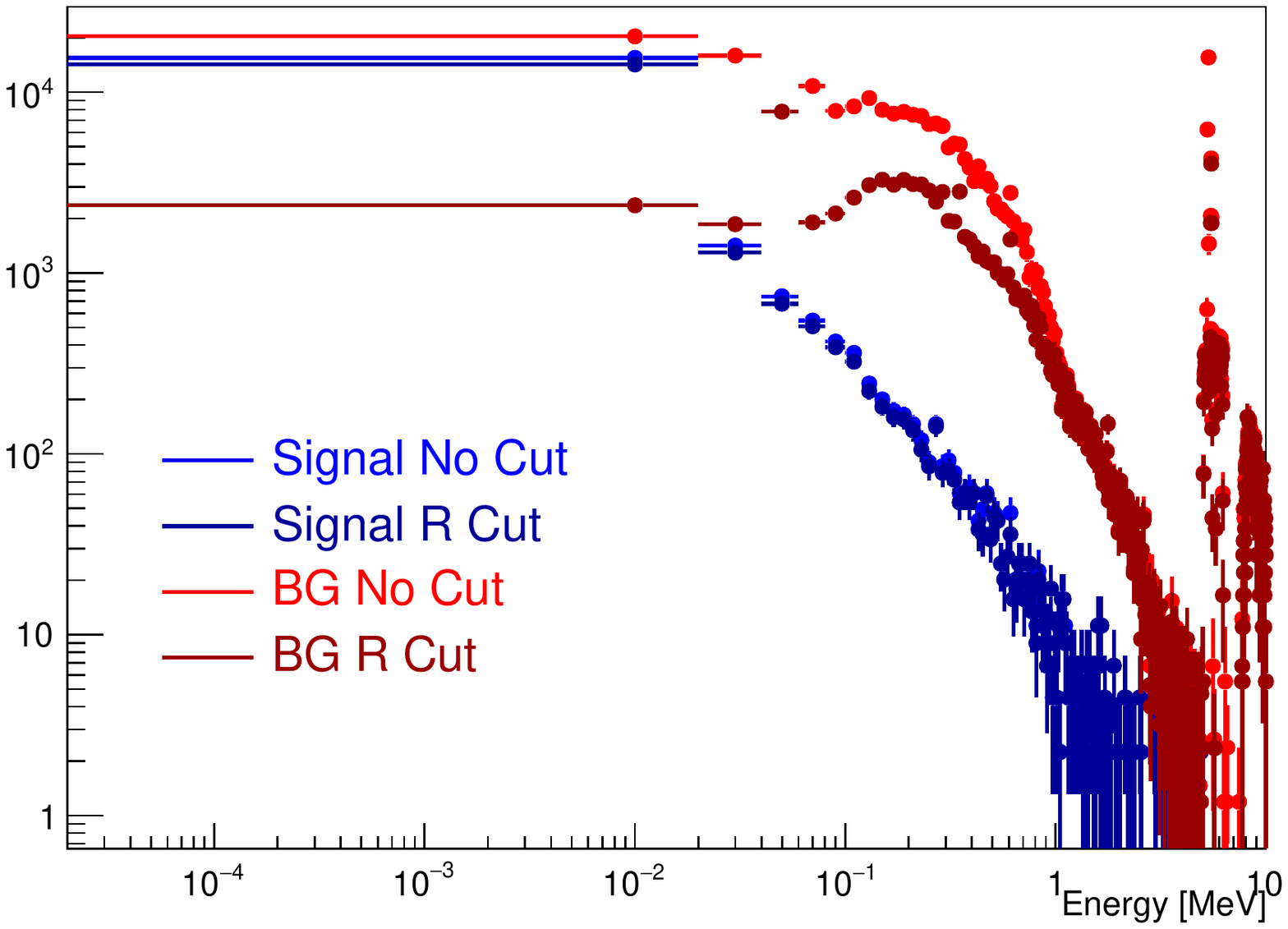}}
\subfigure[\label{fig:cut2}]{\includegraphics[height=6.5 cm]{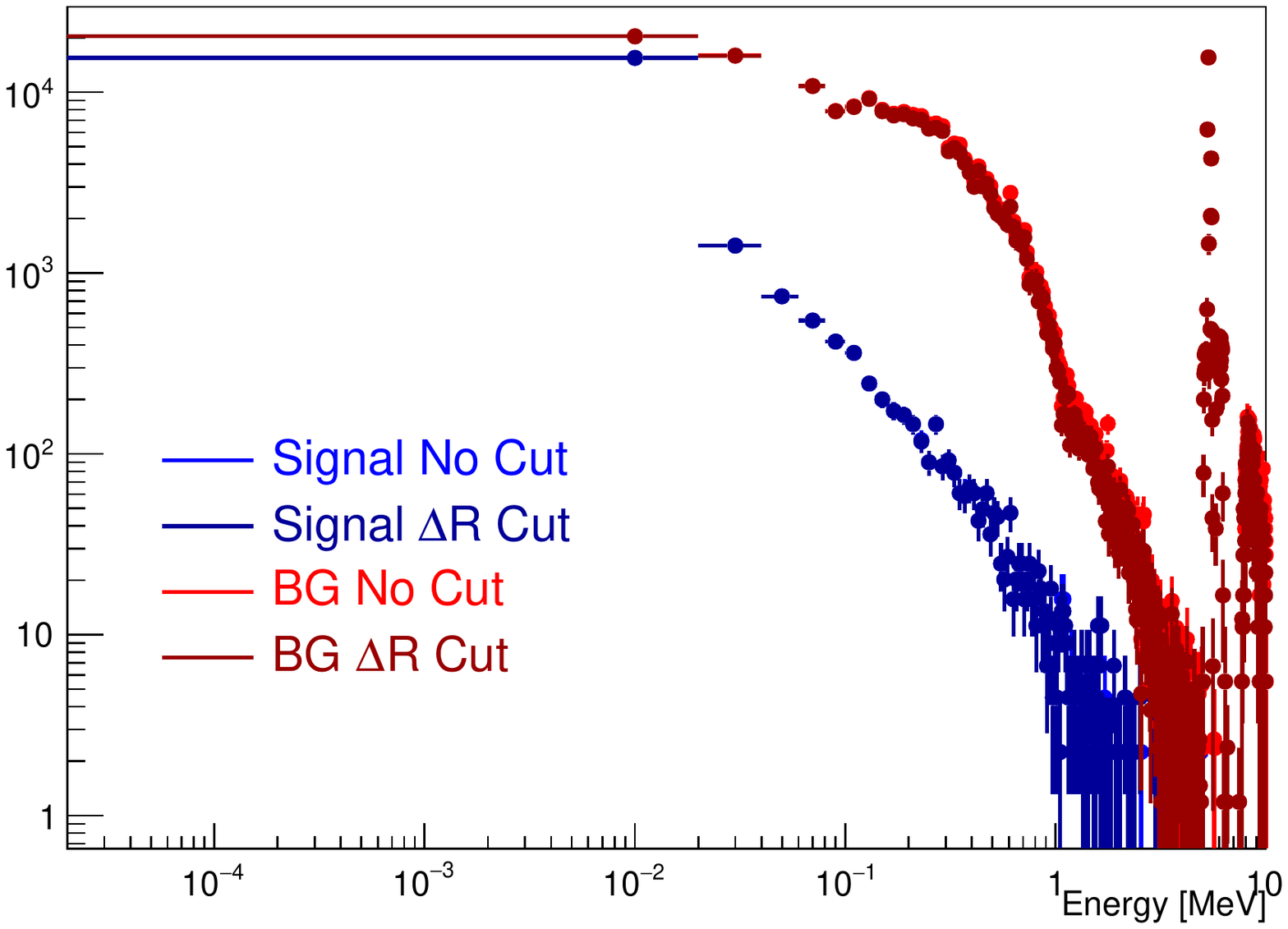}}
\subfigure[\label{fig:cut3}]{\includegraphics[height=6.5 cm]{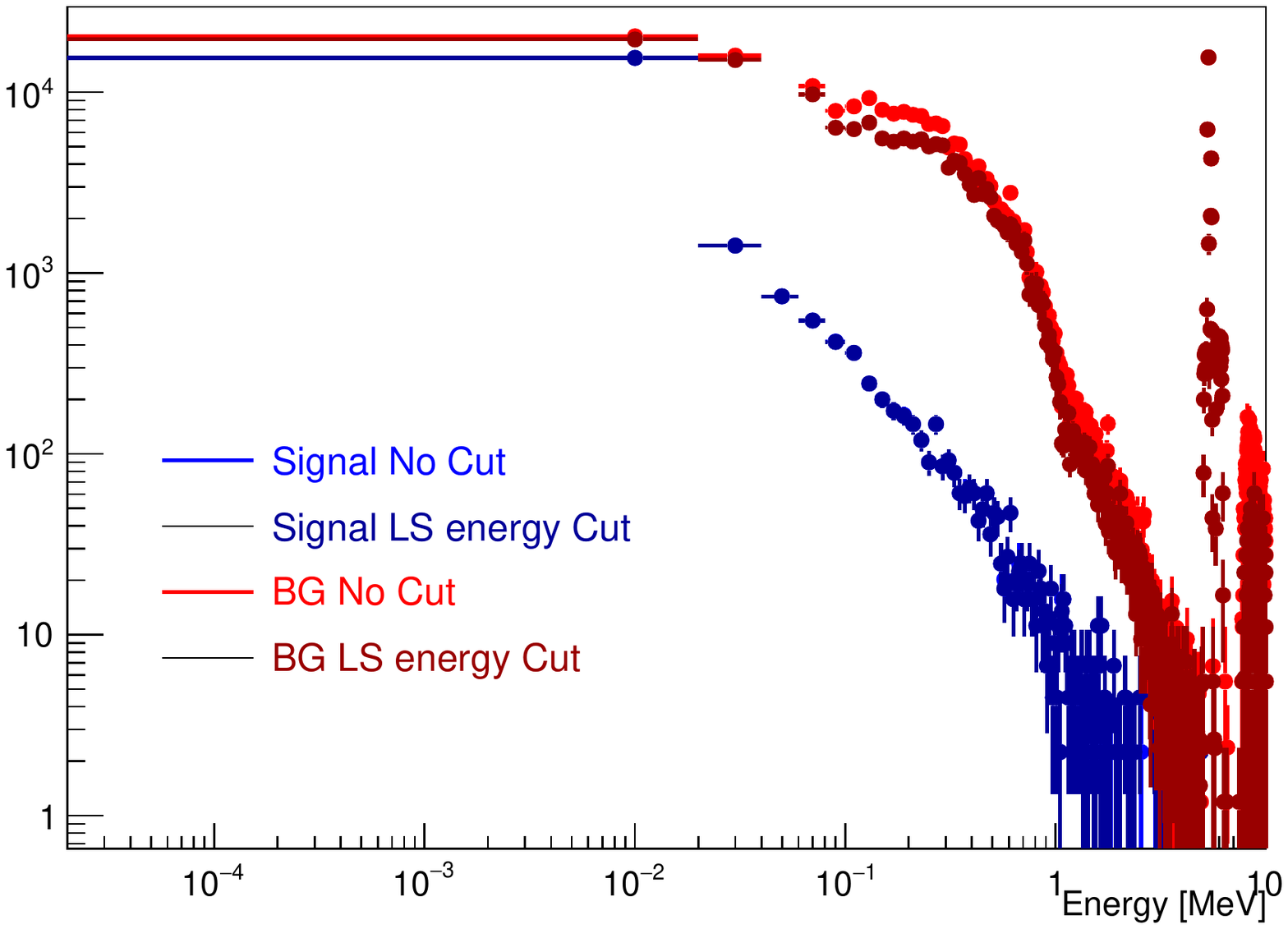}}
\caption{{\it The impact of the cut in radial position~\subref{fig:spectra1}, on $\Delta$R\subref{fig:spectra2} and on the energy deposited in the liquid scintillator\subref{fig:spectra3} are shown for signal and background, without any energy selection.}}
\label{fig:cuts}
\end{center}
\end{figure}

For the signal selection and background reduction the basic cuts established in the framework of the search for $\beta\beta0\nu$ were applied. Full details to justify the applied cuts listed here below, including the different signal and background distributions, can be found in Ref.~\cite{Meregaglia:2017nhx}. The impact of the different cuts is also shown for signal and background without any energy selection in Fig.~\ref{fig:cuts}.

\begin{figure} [p]
\begin{center}
\subfigure[\label{fig:spectra1}]{\includegraphics[height=6.5 cm]{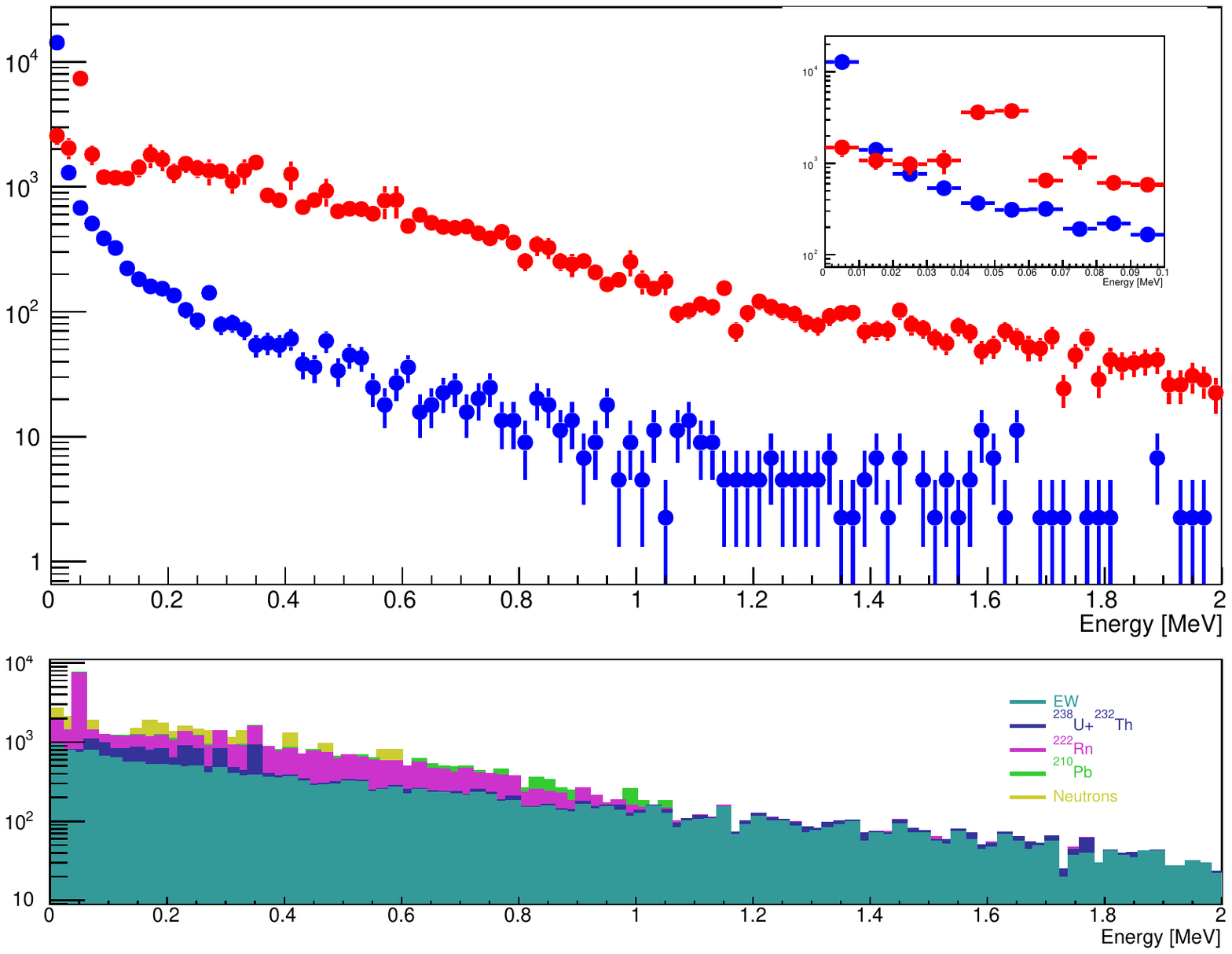}}
\subfigure[\label{fig:spectra2}]{\includegraphics[height=6.5 cm]{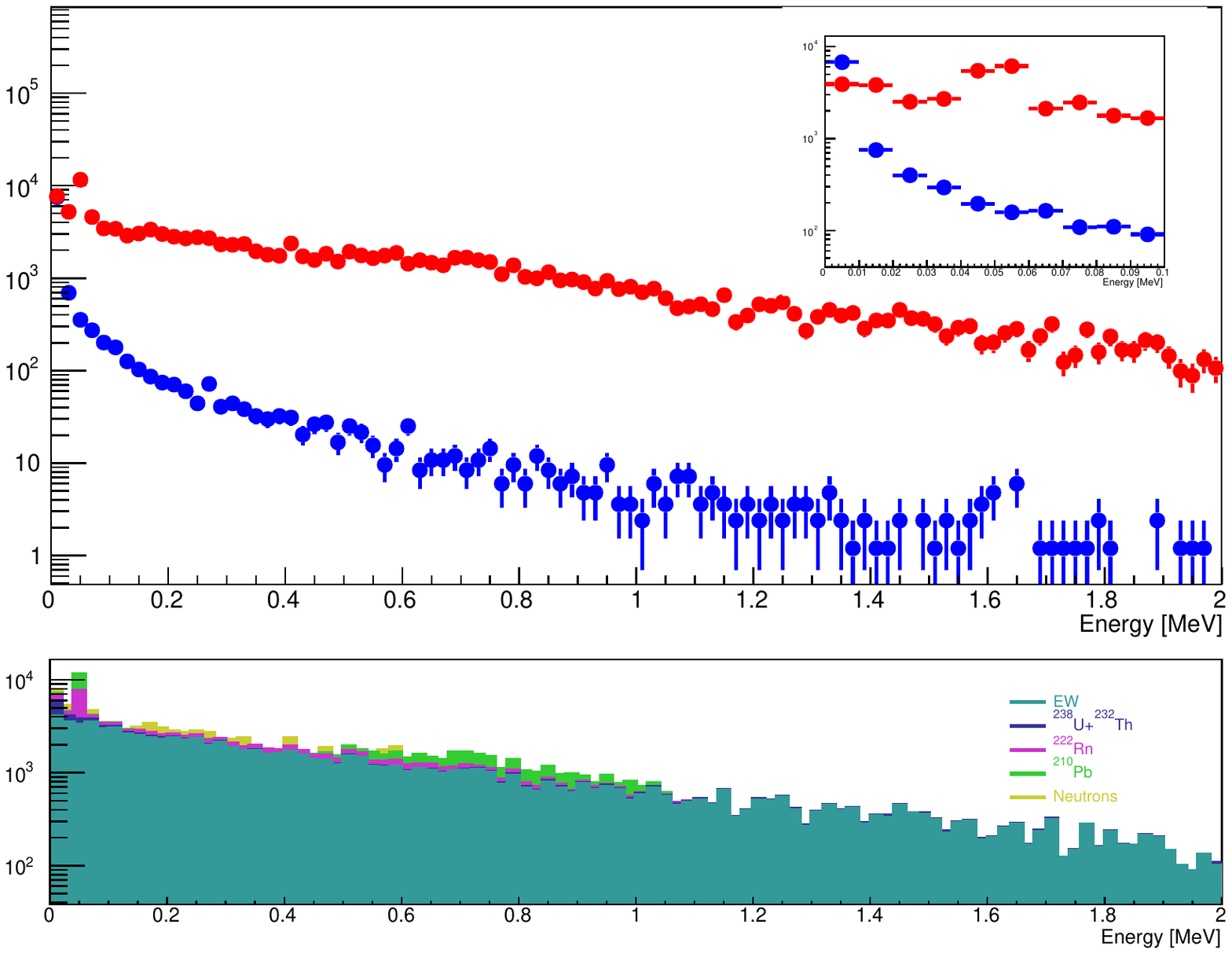}}
\subfigure[\label{fig:spectra3}]{\includegraphics[height=6.5 cm]{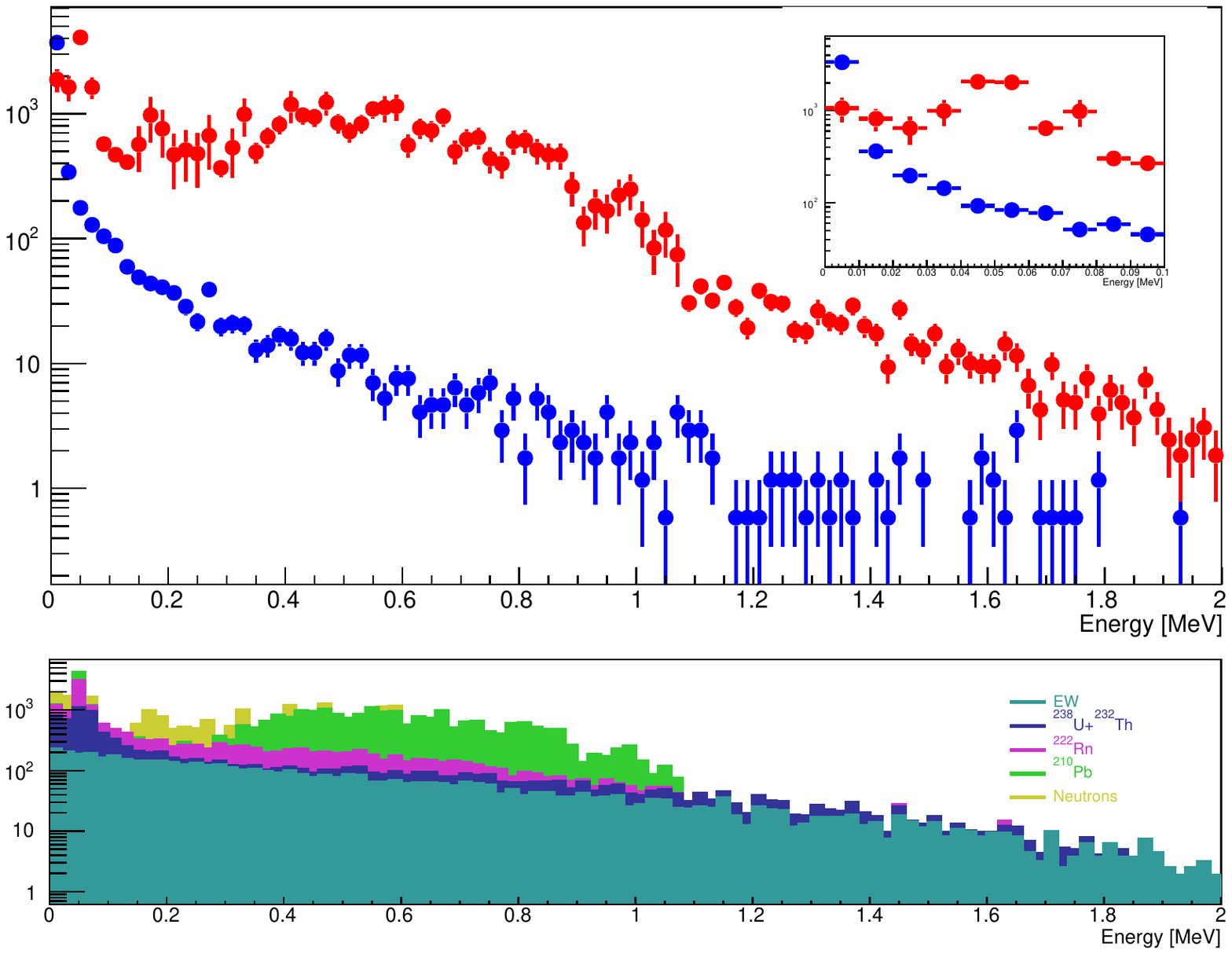}}
\caption{{\it In the upper part of each figure the spectra of reconstructed energy for signal (blue) and background (red) are shown at 40~bar for~\subref{fig:spectra1} Xe,~\subref{fig:spectra2} CF$_4$ and ~\subref{fig:spectra3} Ar. On each figure a zoomed image of the low energy part is also shown. In the bottom part of each figure a breakdown of the different background components is presented.}}
\label{fig:spectra}
\end{center}
\end{figure}

\begin{itemize}
\item A radial cut rejecting events fully contained in the outermost cm of the gas volume. Such a cut, based on the waveform parameters such as the signal risetime, allows to reject alpha events.
\item A cut of 5~cm on the maximal difference between radial deposited energy positions (called $\Delta R$). This cut allows to reject multi-Compton events due to gammas.
\item A cut on the maximal energy deposited of 200~keV in the liquid scintillator. Geant4 simulation showed that a threshold of 200~keV allows for an easy detection of the emitted light with wavelength shifting fibres without adding complexity to the detector and introducing materials which could degrade the detector radio-purity. This cut allows to reject events generated outside the active volume.
\end{itemize}
In the MC a smearing of the energy deposited in the gas was applied assuming a resolution of 1\% FWHM at 2.458 MeV and rescaling it as $\sqrt E$. Such a value is the resolution goal for the proposed detector in order to observe neutrinoless double beta decays~\cite{Meregaglia:2017nhx}. The R2D2 R\&D has obtained promising energy resolution results in argon~\cite{Bouet:2020lbp} and recently in xenon and previous experiment on gas TPC showed that such a resolution can be achieved up to more than 40~bar~\cite{Bolotnikov}.\\
 The obtained results were used to optmize the detector gas pressure and the ROI used for the NMM signal search using as figure of merit the ratio $S/\sqrt{B+S}$ where $S$ represents the signal and $B$ the background. Indeed the application of the described cuts allowed to reach an almost zero background in a region of energy around 2.458~MeV, however the low energy background, mostly due to gammas depositing less than 400~keV in the detector, is not fully rejected.\\  
To maximise the signal, a target with a large number of electrons should be favoured: in this work three different gases were considered namely  CF$_4$, Ar and Xe. The choice is based on the validated performances of SPC detectors with the proposed gases (R\&D for $\beta\beta0\nu$ and previous experiment MUNU for the search of NMM) which would allow to perform an experiment aiming at the NMM measurement with a limited R\&D. The three gases have a large number of electrons, which is a critical condition to have an as large as possible integrated effective cross section of the searched signal.\\ 
A scan in pressure between 5 and 40~bar was performed to study the $S/\sqrt{B+S}$ ratio and define the optimal working pressure for the detector. A high pressure implies a high signal rate however it also results in a larger containment of the background events, therefore there is no obvious choice a priori. The scan in pressure is limited at 40~bar in order to be conservative since the energy resolution in xenon is degraded at pressures above 60~bar~\cite{Bolotnikov}. In addition at high pressure the operation of SPC detectors becomes more complicated in terms of gas purity requirements, and on the high voltage needed on the central anode~\cite{Bouet:2022kav}.\\
The obtained results, applying an ROI of 0 - 2 MeV are summarized in Tab.~\ref{tab:BG} and the corresponding energy spectra at 40~bar are shown in Fig.~\ref{fig:spectra} whereas the signal was computed for a NMM of  $5\times10^{-11} \mu_B$. It is clear that no matter which gas is used, working at high pressure maximises the $S/\sqrt{B+S}$ ratio.
\begin{table}[tp]
\begin{center}
\begin{footnotesize}
\begin{tabular}{|c|ccc|ccc|ccc|}
\hline
gas& \multicolumn{3}{|c|}{Xe} &  \multicolumn{3}{|c|}{CF$_4$}   &  \multicolumn{3}{|c|}{Ar} \\
pressure (bar) &40  & 10  & 5   &  40  & 10  & 5  & 40  & 10  & 5   \\
\hline
Signal for $\mu_\nu = 5\times10^{-11} \mu_B$  &  19942 & 4961& 2473 & 10566 & 2635& 1304  &  5194 & 1273& 623\\
Signal for $\mu_\nu = 2.9\times10^{-11} \mu_B$  &  6708 & 1669 & 832 &  3554 & 886 & 439 &  1747 & 428 & 210\\
Signal for $\mu_\nu = 1.4\times10^{-11} \mu_B$  &  1563  & 389& 194  &  828  & 207& 102 &  407  & 100& 49\\
\hline
Electro-weak signal & 20166 & 4865& 2209  & 10911 & 2391& 1021& 5173 & 997& 416\\
Acrylic vessel &  3925 & 5030& 4926 & 4209 & 4074& 3242 &  4165 & 3527& 3106\\
Copper LS vessel &  685 & 548& 325 &  575 & 131& 77 &  236 & 100& 42 \\

$^{222}$Rn in LS& 1649 & 1048 & 687& 588 & 294 & 211& 512 & 160 & 74\\
$^{222}$Rn in gas& 19449 & 4789& 2376& 11660 & 2970& 1376& 5764 & 1352& 585\\
$^{210}$Pb on copper vessel surface& 1561 & 16365& 26808& 11452 & 37187& 39725& 17764 & 35593& 37708\\

Neutrons& 7108 & 4442& 3110& 6220 & 1999& 1332& 4000 & 1332& 888\\
\hline
Total background &54543& 37087& 40441&45615& 49046& 46984&37614& 43061& 42819\\
\hline
$S/\sqrt{B+S}$ (at  $\mu_\nu = 5\times10^{-11} \mu_B$) &73.1& 24.2& 11.9&44.6& 11.6& 5.9 &25.1& 6.0& 3.0\\

\hline
\end{tabular}
\end{footnotesize}
\caption{{\it  Number of expected events for signal and background in one year for different gases and pressures in an ROI of 0 - 2~MeV for an anti-neutrino flux of $10^{13}$ cm$^{-2}$ s$^{-1}$.}}
\label{tab:BG}
\end{center}
\end{table}%

\begin{figure} [t]
\begin{center}
\subfigure[\label{fig:scanL}]{\includegraphics[height=5 cm]{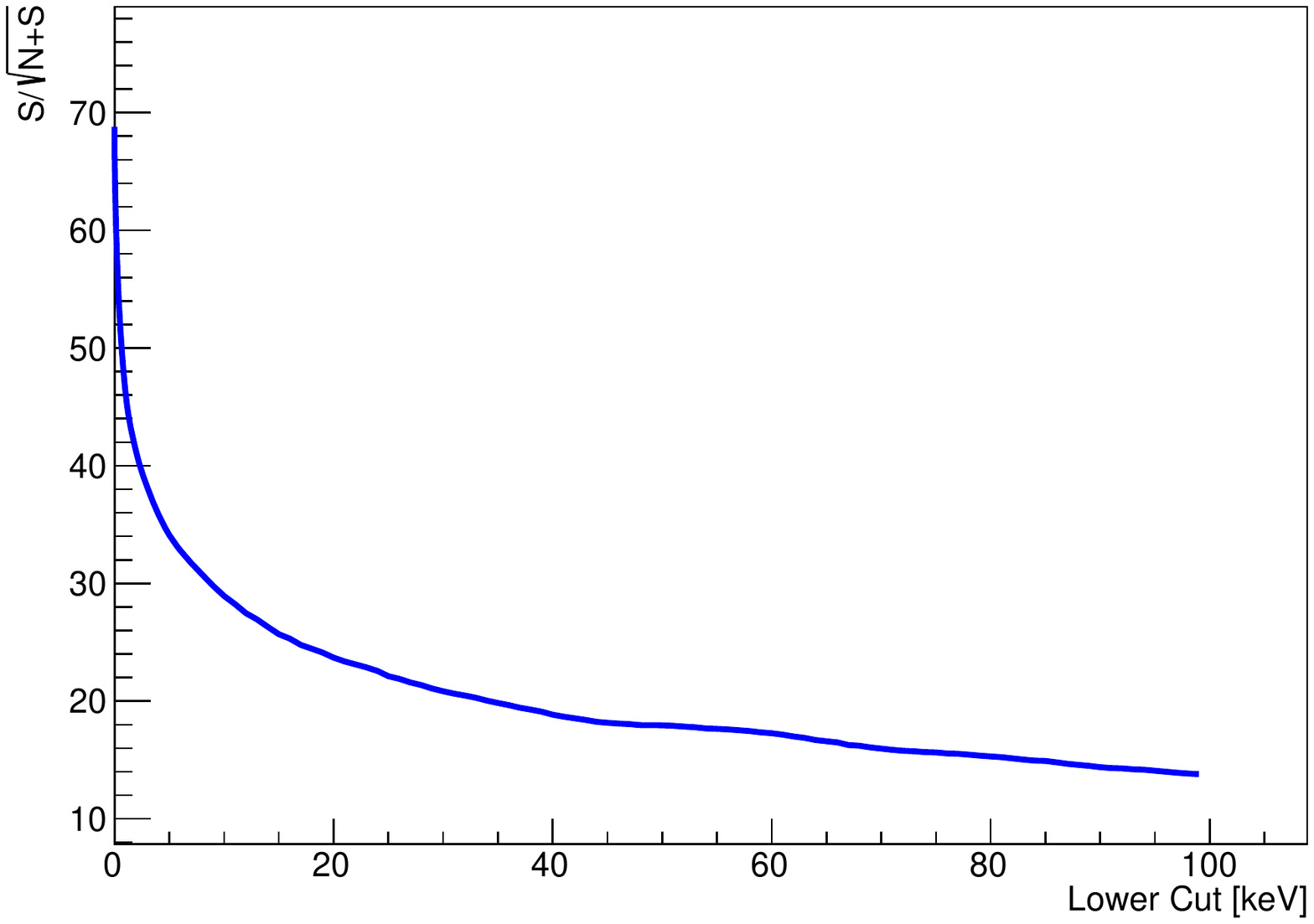}}
\subfigure[\label{fig:scanU}]{\includegraphics[height=5 cm]{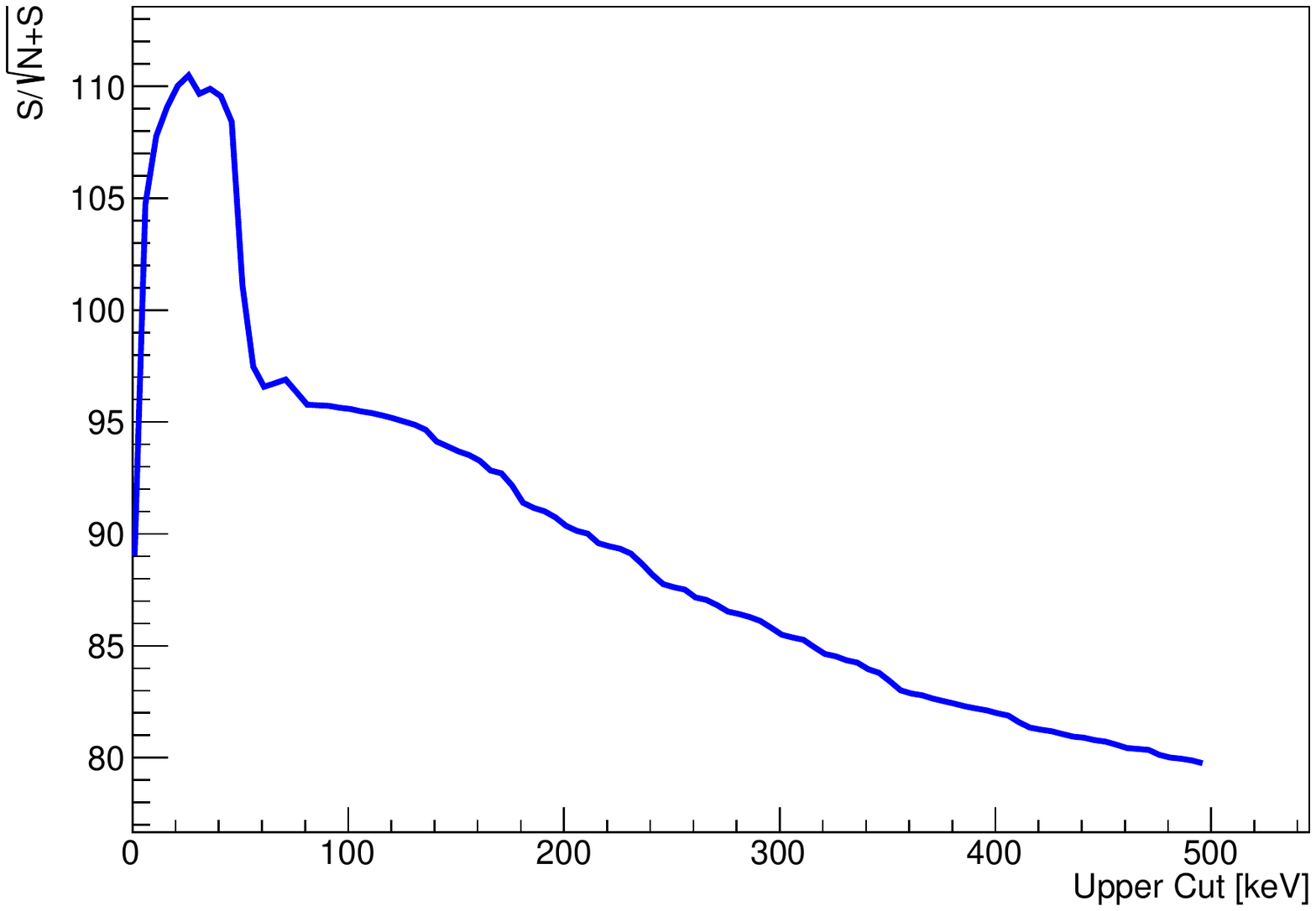}}
\caption{{\it $S/\sqrt{B+S}$ scan in xenon as a function of lower threshold~\subref{fig:scanL} fixing an upper threshold of 2~MeV, and as a function of upper threshold~\subref{fig:scanU} fixing a lower threshold at zero. }}
\label{fig:scan}
\end{center}
\end{figure}

At the selected pressure of 40~bar, a scan of the lower energy threshold of the ROI was performed between 0 and 100~keV in steps of 1~keV fixing the upper threshold at 2~MeV. The obtained $S/\sqrt{B+S}$ for the three gases under study decreases monotonically while increasing the threshold. The same procedure was applied to optimize the higher energy threshold scanning a range between 1~keV and 500~keV with steps of 5 keV fixing the lower threshold at zero. In this case the threshold dependence is less trivial to describe: in the case of xenon we observe  a peak in the $S/\sqrt{B+S}$ distribution at about 30~keV, the distribution starts dropping at about 45~keV and then it decreases monotonically as the upper threshold increases. The results of the threshold scans for xenon can be seen graphically in Fig.~\ref{fig:scan}.\\
In the sensitivity study an ROI of 0 - 45~keV has been assumed, whereas the lower threshold will actually be the experimental detector threshold. This is in principle very low: a threshold at the level of a single ionisation electron of about 17~eV has been demonstrated~\cite{Gerbier:2013vta,Gerbier:2014jwa}, which is a crucial advantage for the proposed detector technology. Stable runs were taken in Ar with the proposed detector with a trigger threshold on the reconstructed charge which corresponds to 10~eV deposited energy in the detector, and an analysis threshold corresponding to 40~eV. Although the detector is capable of observing single electron signal, in a physics run the threshold should be higher in order to avoid a too high noise. However, assuming to ask for a signal corresponding to few electrons, therefore well above the single electron background, and taking into account that the single electron value changes slightly between the proposed gases, it is reasonable to imagine that the detector could be operated in stable condition with a threshold below 200 eV with any gas. Applying such a threshold the reported results stays almost unchanged. 

\section{Event rate}
The rate of events reaching the detector deserves a dedicated study since even events that are rejected by the analysis cuts, produce electrons that are drifted towards the central anode. A problem could be the possible pile-up with good events affecting the reconstruction. Considering a typical drift time of the order of tens to hundreds of $\mu$s we can safely allow for a rate of few tens of Hz.\\
The different backgrounds were considered and the expected rates were computed for the different gases at 40~bar as reported in Tab.~\ref{tab:rate}. The event rate expected after the application of the selection cuts are also reported for completeness although they are not relevant for the pile-up computation. The total rate is at the level of $2 - 3 \times 10^{-2}$~Hz depending on the gas, which does not represent a problem.\\
Another source, which was not considered as background, since easily rejected by the LS, but which has an impact on the trigger rate, are the cosmic muons. The expected rate, assuming 3~m of concrete overburden as done for fast neutrons, is about 20~Hz. This is by far the largest contribution to the trigger rate and, although it does not represent a showstopper for the experiment, a larger overburden would be desirable. Note that the rate decreases quite rapidly increasing the overburden and a factor of 10 is achieved with 3~m of concrete. Locating the detector at ILL where STEREO took data would reduce the cosmic muon rate to about 2~Hz~\cite{Allemandou:2018vwb}.\\
\begin{table}[tp]
\begin{center}
\begin{scriptsize}
\begin{tabular}{|c|c|c|c|c|c|c|}
\hline
Background & \multicolumn{6}{|c|}{Rate (Hz)}\\
 source& \multicolumn{2}{|c|}{Xe}& \multicolumn{2}{|c|}{Ar}& \multicolumn{2}{|c|}{CF$_4$}\\
& No selection & Selection cuts& No selection & Selection cuts& No selection & Selection cuts\\
\hline
Electro-weak signal &  $7.7\times 10^{-4}$   &  $5.9\times 10^{-5}$  &  $1.9\times 10^{-4}$  & $5.8\times 10^{-5}$  &  $4.0\times 10^{-4}$  &  $5.7\times 10^{-5}$ \\
Acrylic vessel &   $6.0\times 10^{-4}$  &  $1.0\times 10^{-6}$  &$4.6\times 10^{-4}$  & $3.8\times 10^{-5}$  &  $5.1\times 10^{-4}$ &  $7.5\times 10^{-5}$ \\
Copper LS vessel &  $4.7\times 10^{-4}$  & $1.4\times 10^{-7}$  & $1.4\times 10^{-4}$  &$2.6\times 10^{-6}$  & $1.9\times 10^{-4}$ &  $4.9\times 10^{-6}$ \\
$^{222}$Rn in LS& $7.9\times 10^{-4}$  &$9.3\times 10^{-7}$  & $3.3\times 10^{-4}$ & $4.9\times 10^{-6}$  & $5.6\times 10^{-4}$ &  $5.4\times 10^{-6}$ \\
$^{222}$Rn in Xe& $1.7\times 10^{-3}$  & $5.5\times 10^{-5}$   &$5.3\times 10^{-4}$  & $1.9\times 10^{-5}$  & $1.1\times 10^{-3}$  &  $4.2\times 10^{-5}$ \\
$^{210}$Pb on copper vessel surface& $1.0\times 10^{-2}$  & $1.0\times 10^{-6}$  & $1.0\times 10^{-2}$  &$2.1\times 10^{-6}$  & $1.0\times 10^{-2}$ &  $1.5\times 10^{-5}$ \\
Neutrons& $1.6\times 10^{-2}$  &$4.2\times 10^{-5}$  & $7.7\times 10^{-3}$ &$3.5\times 10^{-5}$  & $1.3\times 10^{-2}$ & $2.1\times 10^{-5}$ \\
\hline
Total rate &$3.0\times 10^{-2}$ & $1.6\times 10^{-4}$  & $1.9\times 10^{-2}$ &  $1.6\times 10^{-4}$  & $2.5\times 10^{-2}$ &  $2.2\times 10^{-4}$ \\
\hline
\end{tabular}
\caption{{\it  Event rate in Hz for a detector filled with the studied gases at 40~bar with and without selection cuts. When the selection cuts are applied an ROI of 0-45~keV has been assumed.}}
\end{scriptsize}
\label{tab:rate}
\end{center}
\end{table}%
A final point to be considered is the rate in the liquid scintillator. Since it is used as veto in coincidence with the inner detector the rate should also be small enough to avoid pile up. The expected cosmic muons rate in LS is about 100~Hz which could be further reduced with an additional overburden. Such a rate is low enough to have a reasonable pile-up since the probability of having a muon event in a signal time window of 200~$\mu$s is at the per-cent level. An additional drastic reduction of the pile-up could be obtained if scintillation light of the gas is observed, and preliminary R\&D results demonstrated the feasibility~\cite{Bouet:2022kav} in case of Ar and Xe. In that case the coincidence between the two light signals is done in time windows of hundreds of ns reducing the event rate in the LS well below 1~Hz. If CF$_4$ is used, it is harder to observe such a coincidence, since the light yield for Ar and Xe is at the level of 10$^4$ photons per MeV whereas for CF$_4$ the light yield is only about 300 photons per MeV in the UV region and it goes up to 1200 photons per MeV if the full emission spectra is considered~\cite{Pansky:1994zh}.


\section{Experimental site}
In this work the sensitivity of a SPC detector for the NMM measurement near a nuclear reactor is investigated. A specific nuclear reactor site was not selected and this is the reason why a detailed simulation of the overburden for fast neutron and cosmic muons was not performed. In order to estimate the expected signal as a function of the $\mu_\nu$ value it was assumed the Mueller et al. parameterization~\cite{Mueller:2011nm} for energies above 2~MeV and the Kopeikin one~\cite{Kopeikin:2012zz} below 2~MeV normalized to a flux of $10^{13}$ anti-neutrinos cm$^{-2}$ s$^{-1}$ which was the case for the NUMU experiment at the commercial reactor of Bugey.\\
The results obtained on the sensitivity could have small variations depending on the selected site, however the assumed flux is a reasonable estimate for different sites: the same flux is obtained by CONUS at KBR Brokdorf commercial reactor and STERO at the ILL research reactor.\\
The choice of the site will have a larger impact on the cosmic background rather than on the expected signal.


\section{Sensitivity results}

Once all the backgrounds are evaluated, it is possible to estimate the experimental sensitivity in case no signal is observed. This can be done computing the signal upper limit ($S_{up}$) using a Bayesian upper limit for a Poisson parameter and assuming a uniform prior p.d.f., as explained in Ref.~\cite{Agashe:2014kda}:
\begin{equation}
S_{up}=\frac{1}{2} F^{-1}_{\chi^2} [p,2(n+1)]-b
\end{equation}
where $F^{-1}_{\chi^2}$ is the quantile of the $\chi2$ distribution (inverse of the cumulative distribution), $b$ the expected background and $n$ the number of observed events. The quantity $p$ is defined as:
\begin{equation}
p=1-\alpha(F_{\chi^2}[2b,2(n+1)])
\end{equation}
where $F_{\chi^2}$ is the cumulative $\chi^2$ distribution and $(1-\alpha)$ the confidence level.
Given the large number of expected background events, a Gaussian approximation gives very similar results.\\
Considering the backgrounds discussed in Sec.~\ref{sec:BG}, the limits on $\mu_\nu$ were computed for the three gases and the results are reported in Tab.~\ref{tab:result}.\\
Xenon yields the best results with a limit of $4.3 \times 10^{-12} \mu_B$ whereas the sensitivity which can be reached with CF$_4$  and argon is slightly worse, mostly due to the lower number of electrons implying a lower signal.\\

\begin{table}[tp]
\begin{center}
\begin{tabular}{|c|c|c|c|}
\hline
 &    &   \\
Gas  & 90\% C.L. & 3~$\sigma$ C.L. \\
 &    &    \\
\hline
Ar & $8.5 \times 10^{-12} \mu_B$ &  $1.1 \times 10^{-11} \mu_B$ \\
\hline
Xe  &$4.3 \times 10^{-12} \mu_B$ & $5.8 \times 10^{-12} \mu_B$\\
\hline
CF$_4$ & $6.5 \times 10^{-12} \mu_B$ &$8.8 \times 10^{-12} \mu_B$\\
\hline
\end{tabular}
\caption{{\it Expected sensitivity on NMM for the three gases under study. 
}}
\label{tab:result}
\end{center}
\end{table}%


\section{Conclusions}
 
The NMM measurement is of critical importance since a discovery of a non-zero value would be an unambiguous signal of physics beyond the Standard Model. Indeed a measurement of  a value of the NMM larger than $10^{-19}\mu_B$ would require new physics other than the existence of massive Dirac neutrinos to be explained.\\
A spherical high pressure TPC filled with xenon at 40~bar could slightly improve the current limit obtained by XENON1T collaboration, exploiting reactor anti-neutrinos instead of solar neutrinos, in just one year of data taking. Furthermore the proposed detector would profit from the developments carried out already in the framework of the direct Dark Matter search (NEWS-G experiment) or for the search of neutrinoless double beta decay (R2D2 project) without requiring an extensive R\&D.\\

\acknowledgments
The authors would like to thank Dimitris Papoulias from the University of Ioannina (Greece) for useful discussions.

\bibliographystyle{ieeetr}
\bibliography{references}

\end{document}